\documentstyle[12pt]{article}
\load\scriptsize{\sc}
\begin{document}

\font\bigbf=cmbx12 scaled 2300
\def\go{\mathrel{\raise.3ex\hbox{$>$}\mkern-14mu
             \lower0.6ex\hbox{$\sim$}}}
\def\lo{\mathrel{\raise.3ex\hbox{$<$}\mkern-14mu
             \lower0.6ex\hbox{$\sim$}}}
\newcommand{\etal}{{\it et al.\/}}

\begin{center}
{\bigbf Extreme Astrophysics with Neutron Stars}

\vspace*{1truecm}
{\large\bf A Whitepaper Submitted to the Astro~2010 Decadal Survey Committee}
\end{center}

\vspace*{1.5truecm}

\hspace*{1.9truecm} Dong Lai (Cornell University)\footnote{Contact person:
Department of Astronomy, Cornell University, Ithaca, NY 14853. Email: dong@astro.cornell.edu}

\hspace*{1.9truecm} Marten van Kerkwijk (University of Toronto)

\hspace*{1.9truecm} Jon Arons (UC Berkeley)

\hspace*{1.9truecm} Andrei Beloborodov (Columbia University)

\hspace*{1.9truecm} Edward Brown (Michigan State University)

\hspace*{1.9truecm} James Cordes (Cornell University)

\hspace*{1.9truecm} Alice Harding (Goddard Space Flight Center)

\hspace*{1.9truecm} Vicky Kaspi (McGill University)

\hspace*{1.9truecm} Shri Kulkarni (Caltech)

\hspace*{1.9truecm} Duncan Lorimer (West Virginia University)

\hspace*{1.9truecm} Maura Mclaughlin (West Virginia University)

\hspace*{1.9truecm} Roger Romani (Stanford University)

\hspace*{1.9truecm} Anatoly Spitkovsky (Princeton University)

\hspace*{1.9truecm} Tod Strohmayer (Goddard Space Flight Center)


\newpage

\section{Introduction}

This white paper is directed primarily to the ``Stars and Stellar
Evolution'' panel of the Astro 2010 Decadal Survey. Among the various
questions that pertain to this panel, we identify two that are
particularly important and relevant to this white paper:

\medskip
{\bf 1. What are the property and dynamics of matter and radiation under
extreme astrophysical conditions far beyond the reach of
terrestrial environments?}

\medskip
{\bf 2. What are the final products and their observational
manifestations of massive stars as a function of their initial
conditions (mass, metallicity, rotation and magnetic field)?}

\medskip
While the second question is specific to the area of stellar
evolution, the first question is relevant to all areas of
astrophysics.  Indeed, one of the appealing features of astrophysics
among all sciences is that it allows us to probe
extreme conditions far beyond the reach of terrestrial
laboratories. As an end state of stellar evolution, neutron stars
(NSs) are associated with some of the most exotic phenomena and
environments in the universe since the big bang.  Most uniquely, the
supernuclear density in the interior, the strong magnetic field on the
surface, and the intense radiation field in the magnetosphere,
combined with the strong gravity, provide conditions that cannot be
created in terrestrial laboratories. Understanding various
observations of NSs necessarily entails exploring physics under extreme
conditions.  {\it The diversity of the extreme phenomena related to NSs and
the variety of physics tools needed to understand them are what continue to
make this area of research attractive to young researchers
and general public alike.}

In the following we highlight recent progress in NS astrophysics and
discuss prospects for the next decade.  Due to space limitation, our
choices of topics are necessarily selective\footnote{We are aware of
  several other whitepapers focusing on various specific topics
  related to NSs, such as using radio pulsar timing to test theories
  of gravity and to measure gravitational wave background, using X-ray
  timing and X-ray spectroscopy to constrain nuclear equation of
  state, etc.  Here we focus on
  subjects that are not covered by these other white papers, though
  some overlap is unavoidable. In addition, several related topics,
  such as core-collapse supernova explosion mechanism and
  supernova remnants are left out in this whitepaper.}.
But they clearly show that there are many important,
unsolved problems related to NSs, and there exist strong connections
with many areas of astronomy and fundamental physics. Given the 
observational and theoretical progresses in the last five years, 
we believe that the next decade holds great promise for significant
advances in this field. 
{\it A strong program in theory, numerical simulation and
phenomenological modeling, combined with new
sensitive observations/surveys in radio and X-rays/gamma-rays, 
as well as closer interactions with other physics communities 
(gravitational waves and nuclear astrophysics), are
needed to bring out the future advances.}

\newpage
  

\section{Neutron Star Diversity: Formation and Evolution}

Over the last decade, it has become increasingly clear that even in
isolation, NSs have diverse observational manifestations.  More than
2000 radio pulsars are known today, emitting photons and particles as
they slow down. A peculiar class of pulsars (``RRATs'', see
section 3 below) have been discovered in the last few years.
The enigmatic Soft Gamma Repeaters and Anomalous
X-ray Pulsars have been firmly established as magnetars, NSs powered
by the dissipation of superstrong magnetic fields ($B\go 10^{14}$~G).
In addition, a number of ``clean'', thermally emitting and slowly
rotating NSs have been studied extensively. What are the connections
(if any) between these different populations of NSs?
Inside supernova remnants (SNRs), not just radio pulsars and
magnetars, but also another, diverse and puzzling set of young NSs are
found (sometimes called Compact Central Objects), many with
surprisingly little thermal emission.  Indeed, a major puzzle is that
in many SNRs, no discernable compact remnants are seen at all, despite
detailed studies.  Are these NSs that somehow cooled faster, maybe
because they are more massive, and with fields and spins too slow to
produce magnetospheric emission?  Or are they supernovae that led to
low-mass black holes?

Obviously, it is important to understand how these different
types of NSs are related to each other, and how the collapse of
massive stars may give rise to different flavors of NSs (and black
holes). 

Understanding the physical mechanism of core-collapse supernovae is a
long-standing problem in astrophysics (and a subject which surely
deserves a separate whitepaper). There has been significant
progress over the last decade: The standard neutrino-driven
``delayed'' explosion mechanism has been supplemented to incorporate
the important roles played by various hydrodynamical instabilities.
With the continued improvement of computational
capability, the effects of rotation and magnetic field have now begun to
be addressed, as well as more refined treatment of microphysics,
neutrino transport and general relativity.

The birth properties of NSs provide one of the most important constraints
on SN theory. The inter-connection between the study of young NSs, supernova
theory/simulation and the study of advanced stellar evolution
of massive stars, will continue to be fruitful in the next decade.

Another reason we believe success is likely, is that, as will be clear
below, our physical understanding of the emission processes of the NS
surface and magnetosphere is catching up with the detailed
observations.  Furthermore, we are starting to see glimpses of how the
diverse populations of NSs are related to each other.  For instance,
the nearby, thermally emitting NSs appear likely descendants of the
magnetars -- based on kinematic ages and magnetic fields inferred from
timing studies; some of the central compact objects appear to be
weakly-magnetized NSs -- again based on timing studies.

For the next decade, to understand the diverse populations of
NSs and their origin and evolution, what is needed most would be 
further theoretical works on the properties of different classes 
of NSs (some of which are outlined in the next two sections) 
and higher-sensitivity X-ray and radio observations.
In particular, X-ray observations are crucial to study the faint NSs in
SNRs, and to either find NSs that appear to be missing or show no NSs are
present.


\newpage

\section{Pulsars and Magnetars: 
Probing the Dynamics of Highly Magnetized Relativistic Plasma}

\subsection{Observations}

More than 40 years after the discovery of pulsars, observations
continue to reveal new, surprising behaviors of these enigmatic
objects and raise new questions.  
For example, some of the new
findings over the past five years include: (1) A number of radio
pulsars appear to have magnetic fields comparable to those of
magnetars. Could these pulsars turn into magnetars or vice versa (as
recent observations seem to suggest), or are they intrinsically
different?  (2) Some radio pulsars have been found to exhibit distinct
``on'' and ``off'' states with a long term (weeks to years)
quasi-periodicity. 
What is causing this quasi-regular
behavior and intermittency? 
(3) A new class of neutron stars (dubbed {\it Rotating Radio Transients},
RRATs) has been discovered that is detectable only through isolated radio 
bursts,
with typical burst duration of about 1 second and duty cycle of hours to a day.
Some RRATs appear to have magnetar-like magnetic fields, 
while others have spin-down properties similar to those of regular pulsars.
The origin of this ``single pulse'' behavior is not understood, 
and the discovery raises the possibility that the population of RRATs may be
larger than that of normal radio pulsars.
Searches for more of these different flavors of NSs
(e.g., with {\it Arecibo}, {\it GBT} 
and future {\it SKA}), high-energy observations, and dedicated 
radio timing programs 
are necessary to elucidate their physical nature and to constrain the 
possible evolutionary relationships between them.


Pulsars, of course, have been detected in all
wavelength bands, from radio to gamma-rays.
In particular, the {\it EGRET} discovery of seven gamma-ray pulsars 
in the 1990s left important unsolved problems concerning the origin 
of these gamma-rays.
The recently launched {\it Fermi Gamma-Ray Space Telescope} is finding
many sources, with already more than three dozen gamma-ray pulsars
found in six months.  Intriguingly, about half these are found from
their gamma-ray pulsations alone.  These new discoveries are providing a
major leap in our understanding of pulsar particle acceleration sites
and pair cascades physics and also of the census of NSs in the Galaxy.
Hard X-ray observations in the future (such as {\it NuStar} and the European 
{\it Simbol-X}) will add to these
advances by finding the X-ray pulsations of the new gamma-ray pulsars and 
imaging their associated pulsar wind nebulae (PWN).  
Broadband phase-resolved spectroscopy will help 
identify the mechanisms producing the non-thermal emission. Modeling 
the geometry of the 
PWN can provide a measure of our viewing angle to the pulsar rotation axis,
thereby constraining the pulsar emission geometry.
X-ray polarimetry (e.g., such as would be provided by the recently proposed 
{\it Gravity and Extreme Magnetism SMEX}, {\it GEMS}) would add another puzzle piece that 
light curves and spectra alone cannot provide, as the polarization
degree and position angle
can be used to map out the magnetic field of the emission region.

Observations of magnetars in the last few years continue to reveal
surprises, but also have had some nice confirmations of theory.  An
example of the latter is that the quiescent luminosities of several
magnetars with sufficiently accurate distances are found to be around
$10^{35}{\rm\,erg\,s^{-1}}$, as predicted. A recent surprising
discovery was that apart from soft X-ray photons, the emission from
quiescent magnetars is dominated by hard X-rays, with a spectrum
peaking above 200~keV!  Also, many magnetars have been observed to
exhibit bursts and flares of various strengths and on various
timescales, providing diagnostics of the dynamics of highly magnetized
NS crusts (e.g., the yielding behavior of the solid) and
magnetospheres.  Three giant flares have been detected in three
(different) magnetars, with the December 2004 flare from SGR 1806$-$20
radiating energy exceeding $10^{46}$~erg in a few seconds.  Such giant
flares are thought to be powered by global magnetic field
rearrangement/dissipation in the star (e.g., due to a star quake).
The quasi-periodic oscillations observed during these giant flares may
be the first example of NS seismology.  Yet another recent surprise
was that at the end of the active phase of two (``transient'')
magnetars, the stars switched on as bright radio pulsars on their way
to quiescence, with an unusual, hard spectrum of radio
emission. Future timing and spectral observations at different
wavelength bands are important to elucidate many of emerging
phenomenology of magnetars and their connections with radio pulsars.

\subsection{Theories}

Along with the new observational/phenomenological progresses
of pulsars and magnetars described above, there has been significant
recent advance in our understanding of the physics of NS
magnetospheres, which are responsible for the nonthermal emission from
the stars.  The magnetosphere is dominated by highly magnetized,
relativistic electron-positron pairs produced by electromagnetic
cascades.
As mentioned above, the {\it Fermi} Telescope is providing important 
information on the site of particle acceleration and pair cascades in pulsars.
On the theoretical side, 
the global configuration of the dipolar magnetosphere of a rotating NS
has recently been established by means of numerical simulations.
In particular, the distribution
of electric current density in the magnetosphere was found 
to be inconsistent with previous models of 
plasma flow above the polar caps of pulsars --- these models will be revised
in the coming years. New approaches to modeling $e^\pm$ discharges together 
with state-of-the-art numerical simulations 
hold great promise in this regard.
Solution of this fundamental problem, which depends on the application
of modern kinetic plasma simulation techniques not available in
earlier decades and which will make direct contact with observations
through modeling of the pulsed gamma-ray emission, holds much promise
for creating successful models for the mysterious collective radio
emission. Theoretical modeling of plasma interactions 
in the double pulsar system, and others that may be found
in upcoming surveys, offers a valuable means to probe pulsar magnetospheres 
directly. 

In contrast to ordinary pulsars, magnetars exhibit large temporal
variations in their spindown rates, which indicate that their
magnetospheres are {\it dynamic}.  They are likely twisted by
starquakes --- sudden strong deformations of the crust.  The theory of
twisted magnetospheres,
analogous to solar corona, has been developed in the last few years to
account for various observations of magnetars.  Like the solar corona,
the twisted magnetosphere of a NS contains free energy, which is
gradually dissipated and radiated away. A new electrodynamic theory
for dissipative, resistively untwisting magnetospheres has recently
been developed, which attempts to describe the post-starquake
evolution of the luminosity and spindown rate of magnetars. It is
useful to note that because of the dynamic nature of magnetars, the
combination of spectral and torque information provides more
observational constraints on the current flow in the
magnetosphere than available for ordinary radio pulsars.  Thus the
magnetospheres of magnetars may be more trackable than those of radio
pulsars.  Future progress in understanding magnetar dynamics may
significantly impact the landscape of NS research.
Importantly, the superstrong magnetic fields in the magnetar
magnetosphere offer a unique laboratory for the study of exotic
quantum electrodynamic effects, in a regime that is inaccessible for
other astronomical objects or terrestrial laboratories.

\newpage

\section{Thermally Emitting Neutron Stars: Probing Ultra-Dense Matter}

It has long be recognized that thermal, surface emission from NSs has
the potential to provide invaluable information on the physical
properties and evolution of NSs (equation of state at super-nuclear
densities, superfluidity, cooling history, magnetic field, surface
composition, different populations).  With {\it Chandra} and 
{\it XMM-Newton}, the last decade has seen significant observational
progress, revealing the surface magnetic field geometry of isolated
pulsars with phase-resolved spectroscopy, and constraining the cooling
physics from thermal emission of young NSs in SNRs.
In addition, thermal emission from the seven isolated, radio-quiet NSs as 
well as from NSs in quiescent X-ray binaries has been studied in detail.
Overall, the goal is to measure
temperatures and luminosities, and thus radii, as well as
gravitational redshifts and surface gravities.  The former requires
emission from the whole surface, the latter the presence of spectral
features.  Both, of course, require detailed understanding of the emission
processes.

In the last decade, a major break-through has been the discovery of
absorption features at energies below 1\,keV in six of the seven
nearby, thermally emitting neutron stars, in one NS in a SNR, and in
one of the RRAT sources.  In all these, the features almost certainly
arise from the NS surface, but their identification has, so far, remained
uncertain, mostly because we know neither the composition nor the
state of the surface, with suggestions ranging from gaseous hydrogen
to solid iron (or combinations thereof).  Fortunately, the different
models make clear predictions; e.g., for an electron or proton
cyclotron line one expects much larger variation in energy with
rotational phase than for atomic lines.  These predictions are being
tested with {\em XMM} and {\em Chandra}, but definitive measurements will
require better sensitivity with good energy resolution, especially at
low energies, preferably with polarization information.  This would
require the proposed {\em International X-ray Observatory} ({\em IXO}).

Another breakthrough has been the identification of the cooling
radiation of NSs in quiescent X-ray binaries.  An advantage of these
sources is that they likely have weak magnetic fields and known composition:
pure hydrogen, from accretion and subsequent gravitational settling
(of course, this comes with the disadvantage that they do not have
observable spectral features).  
During the active accretion phase,
reactions in the deep crust heat the interior; when accretion halts
and the source enters quiescence, the deposited heat is radiated from
the surface.  Spectral fitting of this thermal emission can determine
the radiation radius, $(R/D)(1-2GM/Rc^{2})^{-1/2}$, where $R$ and $M$
are the neutron star mass and radius.  Both {\it Chandra} and {\it XMM}
observations have been used to constrain the NS radius in this
way, although the accuracy is not yet sufficient to constrain the EOS.
but with {\em IXO}, we would have
the capability to simultaneously measure both the NS  mass
and radius to a few per cent (with GAIA and/or SIM providing the
required accurate distances to binary companions and/or host globular
clusters).  Having a sample of different masses and radii would in
principle allow one to infer the dense matter EOS by inverting the
$R(M)$ relation.

The cooling of NSs in X-ray transients with long outbursts, such as
KS~1731-260, has also been used to follow the thermal relaxation of
the crust into equilibrium with the core.  This has opened a new probe
into the physics of the NS crust. First, the cooling
timescale is set by the thermal conductivity of the deep inner crust,
while the power-law cooling within the first month is directly related
to the heat flux in the outer layers of the NS crust.  The
inferred NS core temperature, when combined with information
about the long-term accretion history, can be used to constrain the
strength of neutrino emissivity of bulk nuclear matter.  Here, again,
progress is limited by current sensitivities, which would be improved
greatly with {\em IXO}.  The project also requires, however, continued
all-sky monitoring in the X-ray, so that follow-up observations can
map out the cooling within the first two weeks of quiescence. It is
during this early time that observations can reveal the distribution
of heat sources in the shallow outer crust; these shallow heat sources
control the behavior of ``long'' X-ray bursts (helium- and
carbon-powered explosions).

To capitalize on these observations and obtain useful constraints on NSs,
it is important to continue the theoretical study of the physical properties
of the NS surface layers (including atmospheres and crusts). For
example, the NS atmosphere consists of highly nonideal, partially
ionized Coulomb plasma, and the properties of atoms and molecules (which 
are likely responsible for the observed spectral lines) can be
significantly modified by the magnetic field. Radiative transfer
can also be strongly affected by exotic QED processes such as vacuum
birefringence. Future observations with proposed X-ray polarimeters, such as
{\it GEMS}, may directly probe such QED signatures. The physical state of the NS crust
(e.g., crystalline or amorphous) and its transport properties determine the rate
of heat transfer from the interior to the surface. There is already a 
community of nuclear physicists interested in matter at supernuclear density,
with whom continued communication and exchange
will be fruitful.  

Over the next decade, the information obtained from NS thermal
emission discussed here, especially if taken together with what is
obtained from different methods covered in other white papers (such as
using pulsar timing to measure NS masses and rotation rates, using
X-ray timing to measure NS mass to radius ratio, etc.)  has the
potential to place neutron stars in their long-hoped position as
excellent probes of the physics of the densest matter in the universe.

\newpage
\section{Neutron Stars as Sources of Gravitational Waves}

In the coming decade, it is likely that advanced LIGO (and its
European counterpart VIRGO) will achieve sufficient sensitivity to
detect gravitational waves (GWs) from astrophysical sources,
thus opening up a new window onto the universe. Neutron stars will
feature prominently in this window:

The most promising (and conservative) sources of GWs are
coalescing NS/NS binaries. It is even possible that NS/BH will be
detected first in the next few years by the ``enhanced LIGO'' if some
nonstandard (but physically possible) binary evolution scenario 
proves correct.
During the binary inspiral phase, 
the masses of the compact objects (NS or BH) can be accurately
measured from the inspiral waveform. The GW waveform generated in the final
merger (or tidal disruption of the NS by the BH)
may allow us to infer
the NS radius and the stiffness of the nuclear matter,
thus giving a completely new constraint on the nuclear equation of
state. It is may be possible that prior to merger, dynamical tides on
the NS (if it has relatively large radius) lead to detectable
imprints on the GW signals, again shedding light on the NS internal
structure.

Numerical relativity has advanced significantly in the last few years,
and it is now becoming feasible to simulate numerically the whole inspiral
and merger process for many dynamical times. It will be useful to 
incorporate more sophisticated microphysics in the simulations
(e.g., hot nuclear EOS and neutrino transport), as well as to study
and simulate
the effects of magnetic fields. The latter is particularly challenging but 
potentially important, as there has been mounting observational evidence in
the last few years that merging NS/NS or NS/BH binaries
lie at the heart of the central engine for (short/hard) gamma-ray bursts.

Accreting NSs in binary systems may also be detectable sources of GWs
for LIGO. As matter accretes onto the NS, an asymmetric mass distribution 
(quadrupole moment) may develop due to various mechanisms 
(``magnetic mountain'', asymmetric nuclear reactions or overstable 
Rossby waves driven by gravitational radiation reaction). 
There is significant uncertainty in the estimate for the
expected GW strength and much more theoretical work is needed. 
Direct detection of GWs from such sources would
allow us to diagnose both the accretion process and the NS
structure.

A large body of observational and theoretical work over the last decade
has suggested that core-collapse supernovae (SNe) and NS formation involve
highly asymmetric dynamical processes, which are largely hidden
from the electromagnetic window, but can generate GWs. The detection of such GWs from
nearby SNe by LIGO would reveal much about the explosion mechanism.
If rotation is dynamically important (a big ``if'' since there is much 
uncertainty about the angular momentum evolution of pre-SN stars),
much enhanced and qualitatively different GW signals would be generated.

\end{document}